\documentclass[twocolumn, a4paper, 10pt, showpacs, reprint, prb, superscriptaddress, amsmath, amssymb, amsfonts, floatfix, showkeys]{revtex4-2}
\usepackage[vietnamese=nohyphenation]{hyphsubst}
\usepackage[vietnamese,british]{babel}
\usepackage{graphicx,epsfig}
\usepackage{float}
\usepackage{hyperref}
\usepackage{breqn}
\usepackage[utf8]{inputenc}
\usepackage{amsmath}
\usepackage{subfigure}
\usepackage[normalem]{ulem}
\usepackage{xcolor}
\usepackage{soul}
\hypersetup{ 
     colorlinks   = true,
     citecolor    = blue,
     urlcolor     = blue
}
\usepackage{physics}
\usepackage{amssymb}
\usepackage{multirow}

\begin{document}


\title{Charge fluctuations, phonons and superconductivity in multilayer graphene}

\author{Ziyan Li}
\affiliation{$Key \ Laboratory \ of \ Artificial \ Micro\- \ and \ Nano\-structures \ of \ the \ Ministry \ of \ Education \ and \ School \ of \ Physics \ and\ Technology,\ Wuhan \ University, \ Wuhan \ 430072,\ China$}
\affiliation{$Imdea \ Nanoscience,\ C/ Faraday \ 9, \ 28015 \ Madrid,\ Spain$}

\author{Xueheng Kuang}
\affiliation{$Key \ Laboratory \ of \ Artificial \ Micro\- \ and \ Nano\-structures \ of \ the \ Ministry \ of \ Education \ and \ School \ of \ Physics \ and\ Technology,\ Wuhan \ University, \ Wuhan \ 430072,\ China$}
\affiliation{$Imdea \ Nanoscience,\ C/ Faraday \ 9, \ 28015 \ Madrid,\ Spain$}

\author{Alejandro Jimeno-Pozo}
\affiliation{$Imdea \ Nanoscience,\ C/ Faraday \ 9, \ 28015 \ Madrid,\ Spain$}

\author{H\'ector Sainz-Cruz}
\affiliation{$Imdea \ Nanoscience,\ C/ Faraday \ 9, \ 28015 \ Madrid,\ Spain$}

\author{\\Zhen Zhan}
\email{zhenzhanh@gmail.com}
\affiliation{$Imdea \ Nanoscience,\ C/ Faraday \ 9, \ 28015 \ Madrid,\ Spain$}
\affiliation{$Key \ Laboratory \ of \ Artificial \ Micro\- \ and \ Nano\-structures \ of \ the \ Ministry \ of \ Education \ and \ School \ of \ Physics \ and\ Technology,\ Wuhan \ University, \ Wuhan \ 430072,\ China$}

\author{Shengjun Yuan}
\affiliation{$Key \ Laboratory \ of \ Artificial \ Micro\- \ and \ Nano\-structures \ of \ the \ Ministry \ of \ Education \ and \ School \ of \ Physics \ and\ Technology,\ Wuhan \ University, \ Wuhan \ 430072,\ China$}
\affiliation{$Wuhan \ Institute \ of \ Quantum \ Technology,\ Wuhan \ 430206,\ China$}

\author{Francisco Guinea}
\affiliation{$Imdea \ Nanoscience,\ C/ Faraday \ 9, \ 28015 \ Madrid,\ Spain$}
\affiliation{$Donostia \ International \ Physics \ Center,\ Paseo \ Manuel \ de \ Lardizabal \ 4, \ 20018 \ San \ Sebastian,\ Spain$}

\begin{abstract}
Motivated by the recent experimental detection of superconductivity in Bernal bilayer (AB) and rhombohedral trilayer (ABC) graphene, we study the emergence of superconductivity in multilayer graphene based on a Kohn-Luttinger (KL)-like mechanism in which the pairing glue is the screened Coulomb interaction.\ 
We find that electronic interactions alone can drive superconductivity in AB bilayer graphene and ABC trilayer graphene with the critical temperatures in good agreement with the experimentally observed ones, allowing us to further predict superconductivity from electronic interactions in Bernal ABA trilayer and ABAB tetralayer and rhombohedral ABCA tetralayer graphene.\ By comparing the critical temperatures ($T_c$) of these five non-twisted graphene stacks, we find that the ABC trilayer graphene possesses the highest $T_c\sim100$ mK.\ After considering the enhancement of superconductivity due to Ising spin-orbit coupling, we observe that the AB bilayer graphene has the largest enhancement in the critical temperature, increasing from 23 mK to 143 mK.\ The superconducting behaviours in these non-twisted graphene stacks could be explained by the order parameters (OPs).\ The OPs of Bernal stacks preserve intravalley $C_3$ symmetry, whereas rhombohedral stacks break it.\ In all stacks, the OPs have zeroes and change sign between valleys, which means that these multilayers of graphene are nodal spin-triplet superconductors.\ Moreover, dressing the purely electronic interaction with acoustic phonons, we observe minor changes of the critical temperatures in these five stacks.\ We adopt the KL-like mechanism to investigate the tendency of superconductivity in multilayer graphene without fitting parameters, which could provide guidance to future experiments exploring superconductivity in non-twisted graphene.  

\end{abstract}

\maketitle

\section{\label{sec:level1}Introduction}

Recent experiments on Bernal bilayer (AB) and rhombohedral trilayer (ABC) graphene reveal cascades of correlated phases and superconductivity, thus providing a platform to study correlation effects in ultra-clean graphene-based systems~\cite{Zhou2021HalfMetRTG,Zhou2021SuperRTG,Zhou2022SCBG,barrera2022cascade,seiler2022quantum,zhang2022spin,Lin23Spontaneous,holleis2023ising}.\ These non-twisted stacks show plenty of advantages, for instance, they are common allotropes of graphene and are quite stable, in contrast to twisted stacks, which suffer angle disorder~\cite{uri2020mapping} and strains~\cite{kazmierczak2021strain}, require careful fabrication techniques to control the twist angles and yield superconductivity that depends sensitively on the angle~\cite{lau2022reproducibility}.

Ample evidences show that the superconductivity detected in these non-twisted multilayer graphene is unconventional, for a review see Refs.~\cite{pantaleon23SC,pangburn2022superconductivity}.\ Up to now, there is still no sufficient experimental evidence for the mechanism and pairing symmetry of these systems.\
However, both AB and ABC graphene show large Pauli limit violations, evidence of spin-triplet pairing, and superconductivity next to flavour symmetry breaking transitions \cite{Zhou2021SuperRTG,Zhou2022SCBG,zhang2022spin}.\ Strikingly, an in-plane magnetic field promotes a spin-triplet superconducting state in AB graphene~\cite{Zhou2022SCBG}, and so does Ising spin-orbit coupling (SOC)~\cite{zhang2022spin,holleis2023ising}.\ These phenomena point to a purely electronic pairing mechanism.\ 

Previous theoretical works have proposed three main candidates for the pairing in these all-carbon systems:\ electron-phonon coupling~\cite{CWSS22,CWSS22b,chou2022enhanced,Chou2021AcousticRTG};\ flavour fluctuations in proximity to symmetry broken phases~\cite{SR22,DCL22,Szabo2022RTG,ZL21,Chatterjee2022,ZV21}; and the Coulomb interaction, either the screened long-range potential~\cite{PhysRevB.107.L161106,Cea2022SCKLRTG,PhysRevLett.130.146001,Lu2022RG_RTG,PhysRevLett.127.247001, Wagner2023Bernal} or the short-range interaction~\cite{C22,Dai2022QMonteCarlo,DHZLM21}.\ Several theoretical works have advanced the proposal that the Coulomb interaction suffices to drive superconductivity in AB and ABC graphene.\ In particular, Ref.~\cite{PhysRevB.107.L161106} performs a Random Phase Approximation (RPA), Kohn-Luttinger-like analysis and shows that the screened long-range Coulomb interaction accounts for the critical temperatures obtained in experiments.\ Furthermore, it captures the enhancement of superconductivity due to Ising SOC induced by proximity to a substrate and also predicts the form of the superconducting order parameters for AB and ABC graphene, which reveal nodal spin-triplet superconductivity when spin-orbit coupling is absent, and nodal Ising superconductivity when it is present.

Superconductivity has not yet been found in Bernal ABA trilayer graphene, Bernal ABAB tetralayer graphene or rhombohedral ABCA tetralayer graphene. A reasonable question is whether superconductivity exists in these stacks.\ Several theoretical works predict that superconductivity indeed exists in the ABCA graphene~\cite{CWSS22b, Ghazaryan2022Multilayers}.\
Here we work in the RPA, Kohn-Luttinger-like framework and use a full-scale tight binding (TB) model to predict superconductivity from electronic interactions in hole-doped Bernal ABA trilayer and ABAB tetralayer and rhombohedral ABCA tetralayer graphene.\ By comparing them to AB bilayer and ABC trilayer graphene, we study the tendency of superconductivity in non-twisted multilayer graphene as a function of the number of layers and their stacking. We also take into account dressing by phonons.\ 

Our results show that all these stacks are superconductors and that the pairing glue is the Coulomb interaction.\ 
ABC trilayer graphene stands out as the stack with the highest critical temperature.\ Furthermore, the Ising SOC increases the $T_c$ of every stack, but it is most beneficial for the AB bilayer.\ 
The superconducting order parameter in Bernal stacks is $C_3$-symmetric and has intravalley extended $s$-wave symmetry.\ In contrast, the order parameter in rhombohedral stacks breaks $C_3$ symmetry and has intravalley $p$-wave-like symmetry.\ Our results for the AB bilayer and ABC trilayer graphene are in excellent agreement with experiments~\cite{Zhou2021SuperRTG,Zhou2022SCBG}.\ Consequently, predictions of superconductivity in ABA, ABAB and ABCA graphene may provide guidance to future experiments.\ The paper is organized as follows, first we describe the TB model, RPA and Kohn-Luttinger-like approach in Sec.\ \ref{sec:level2}.\ Then we present the results for the critical temperatures and superconducting order parameters of the five different graphene stacks in Sec.\ \ref{result}.\ Finally, we discuss and summarize our work.

\begin{figure}[t]
  \centering
  \includegraphics[scale=0.14]{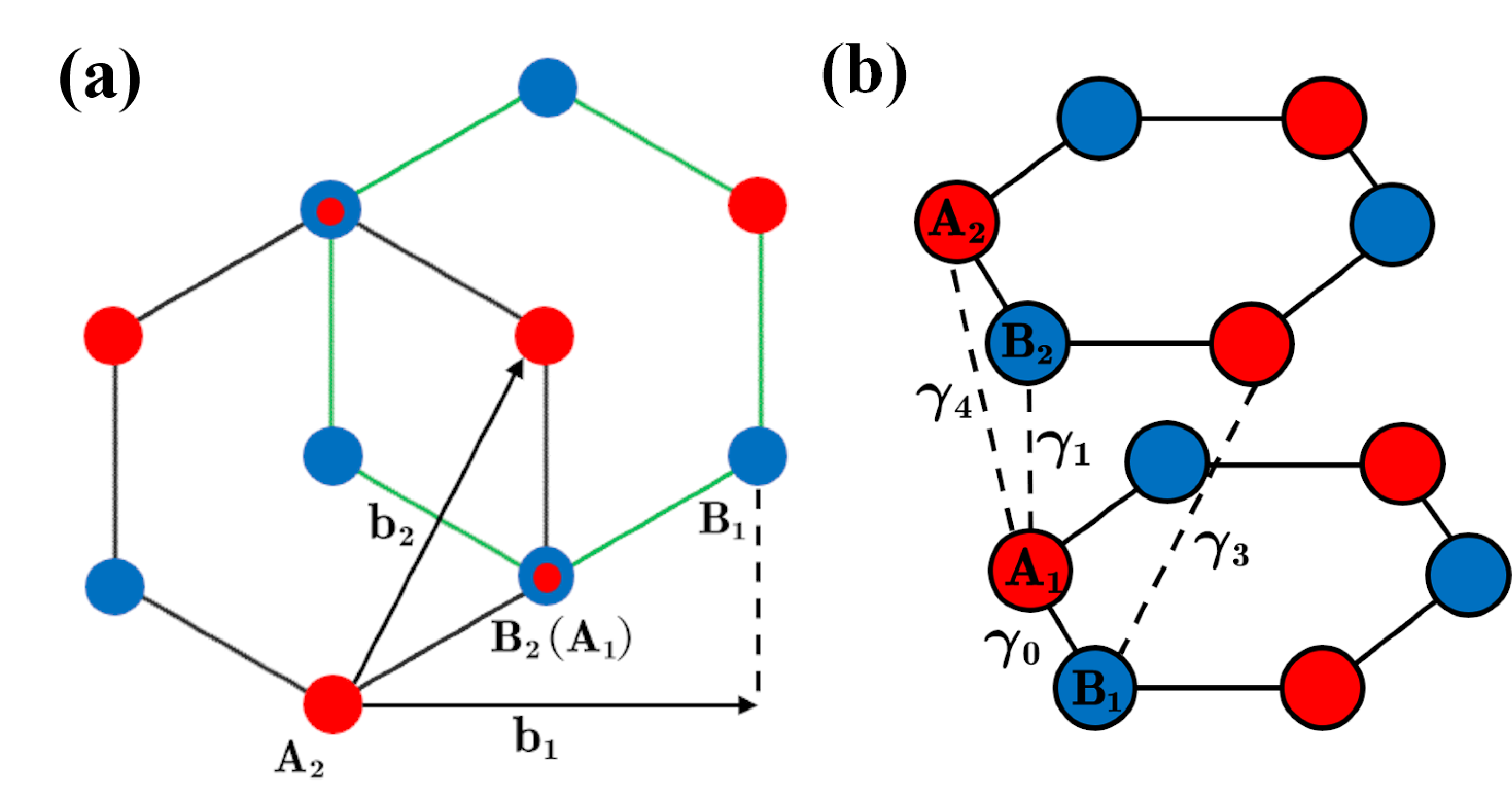}
  \caption{The lattice structure of AB-stacked graphene.\ (a) The 2D crystal structure of AB-stacked graphene, with $\mathbf{b}_1$ and $\mathbf{b}_2$ representing the primitive translation vectors.\ (b) The 3D crystal structure and some representative hopping parameters between atoms.}
  \label{fig:1}
\end{figure}

\begin{center}
\begin{table*}[t]
     \centering
     \caption{The hopping parameters and on-site energies in multilayer graphene systems \cite{wu2015detection, mccann2013electronic}. }
     \setlength{\tabcolsep}{5mm}{
     \begin{tabular}{||c c c c c c c c||} 
         \hline
        Parameters (eV) & $\gamma_0$ & $\gamma_1$ & $\gamma_2$ & $\gamma_3$ & $\gamma_4$ & $\gamma_5$ & $\delta$    \\  [0.3ex]
         \hline\hline
         AB bilayer & 3.16 & 0.381 & — & 0.380 & 0.140 & — & 0.022 \\ 

         ABC/ABA trilayer  & 3.10 & 0.370 & -0.032 & 0.300 & 0.040 & 0.050 & 0.040 \\

         ABCA/ABAB surface bilayer  & 3.18 & 0.390 & -0.012 & 0.300 & 0.040 & 0.020 & 0.050 \\

          ABCA/ABAB bulk bilayer  & 3.18 & 0.385 & -0.012 & 0.250 & 0.030 & 0.020 & 0.050\\
         \hline
     \end{tabular}}

     \label{tab:1}
\end{table*}
\end{center}

\section{\label{sec:level2}Numerical methods} 
\subsection{\label{sec:2_A}Tight-binding model}

We employ a TB model to calculate the band structure of multilayer graphene systems. We briefly introduce this model via the AB bilayer graphene case.\ The term AB refers to shifting one of the graphene layers in the direction along one-third of the translation vector $\mathbf{b}_1+\mathbf{b}_2$, as shown in Fig.~\ref{fig:1}(a).\ The lattice structure and representative hopping parameters are shown in Fig.~\ref{fig:1}(b).\ Each unit cell of AB bilayer graphene consists of four atoms, ${A}_i$ and ${B}_i$ ($i=1,2$).\ Taking into account one $p_z$ orbital per atomic site, the TB Hamiltonian of the AB-stacked graphene is a function of the momentum $\mathbf{k}$ in the first Brillouin zone (BZ) \cite{mccann2013electronic}.\ In the basis $\{ A_{1}, B_{1}, A_{2}, B_{2} \}$ it is given by:
\begin{equation}
    \mathcal{H}_{AB}\left( \mathbf{k}\right) = \mqty(-\frac{\Delta_1}{2}+\delta & -\gamma_0 u(\mathbf{k}) & \gamma_4 u^{\ast}(\mathbf{k}) & \gamma_1 \\ -\gamma_0 u^{\ast}(\mathbf{k})  & -\frac{\Delta_1}{2} & -\gamma_3 u(\mathbf{k}) & \gamma_4 u^{\ast}(\mathbf{k}) \\ \gamma_4 u(\mathbf{k}) & -\gamma_3 u^{\ast}(\mathbf{k}) & \frac{\Delta_1}{2} & -\gamma_0 u(\mathbf{k})\\ \gamma_1 & \gamma_4 u(\mathbf{k}) & -\gamma_0 u^{\ast}(\mathbf{k} ) & \frac{\Delta_1}{2}+\delta)
    \label{equ:1}
\end{equation}
where $\gamma_0$ describes the nearest neighbor hopping amplitude between the atoms within the same layer, $\gamma_i$ ($i$=1, 3, 4) refers to the interlayer hopping parameters and $\Delta_1$ represents the interlayer potential difference between two nearest layers, which can be generated by applying a perpendicular electric field to the graphene system.\ $\delta$ is the on-site potential term that only exists at sites $A_1$ and $B_2$, which have neighbors in the adjacent layers.\ The function $u(\mathbf{k})= \mathrm{e} ^ {-\mathrm{i} k_ya/\sqrt{3}}+2\mathrm{e}^{\mathrm{i} k_ya /2\sqrt{3}}\cos(k_x a/2)$ describes the hopping between nearest  carbon atoms, with $a=0.246$ nm and $d_0=0.333$ nm representing the lattice constant and interlayer distance of graphene, respectively.\ All the hopping parameters in different multilayer graphene are tabulated in Table \ref{tab:1}. Note that the parameters $\gamma_2$ and $\gamma_5$ are extra parameters in trilayer and tetralayer graphene, which indicate the next-nearest interlayer interactions between atoms from the first and third layers.\ For more details about the TB parameters, please refer to Refs.\ \cite{mccann2013electronic, aoki2007dependence, wu2015detection, nanda2009strain}.   

\subsection{Screened Coulomb interaction}
In this part, we analyze the effect of metallic gates, electronic interactions and acoustic phonons in the screening of the Coulomb potential by means of the RPA, following the procedure shown in \cite{cea2020band,yuan2011excitation,cea2021coulomb}.\ Firstly, we assume that the bare Coulomb potential in the graphene multilayer is affected by the experimental set-up, in which the multi-layer is commonly placed within two metallic gates.\ Referring to Refs.\ \cite{cea2020band,alonso2017acoustic}, we set this distance to be the same for the top and bottom layers, with a value of $d=40$ nm.\ The dual-gated Coulomb potential has the following expression in momentum space \cite{cea2021coulomb}:
\begin{equation}
    \textcolor{black}{ V_{C}(\mathbf{q})=
      \frac{2\pi e^{2}}{\epsilon\abs{\mathbf{q}}}  \tanh{(\abs{\mathbf{q}} d)} }
    \label{equ:2}
\end{equation}
where $e$ is the electron charge and $\epsilon=2.7$ \cite{AOKI2007123} represents the dielectric constant due to encapsulation of the systems.\ Notably, from our results (details in Appendix \ref{append_eps}), the superconductivity of multilayer graphene depends weakly on the value of $\epsilon$.

\begin{figure}[h]
  \centering
  \includegraphics[scale=0.127]{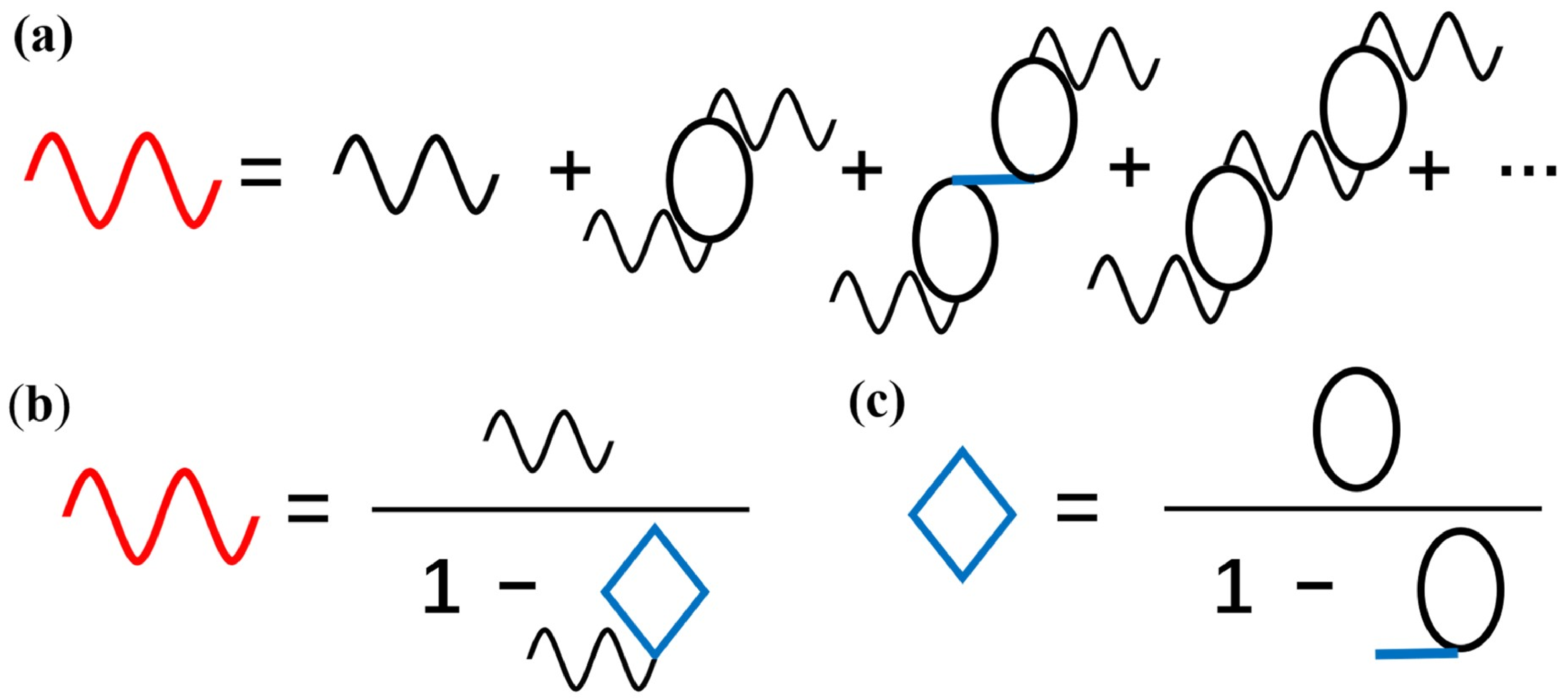}
  \caption{Feynman diagrams describing the screened Coulomb potential.\ (a) The red wavy line represents the total screened potential $V^{tot}_{scr}(\mathbf{q})$ while the black wavy line is dual-gated Coulomb potential $V_{C}(\mathbf{q})$.\ (b) Renormalization of the Coulomb potential, and the blue diamond describes the total screened polarizability $\Pi^{tot}(\mathbf{q})$.\ (c) Renormalization of the bare electronic polarizability $\Pi^{ele}(\mathbf{q})$.\ The straight blue lines stand for the electron-acoustic phonon coupling $V^{ph}(\mathbf{q})$}.
  \label{fig:2}
\end{figure}

\begin{figure*}[t]
  \centering
  \includegraphics[width=\textwidth]{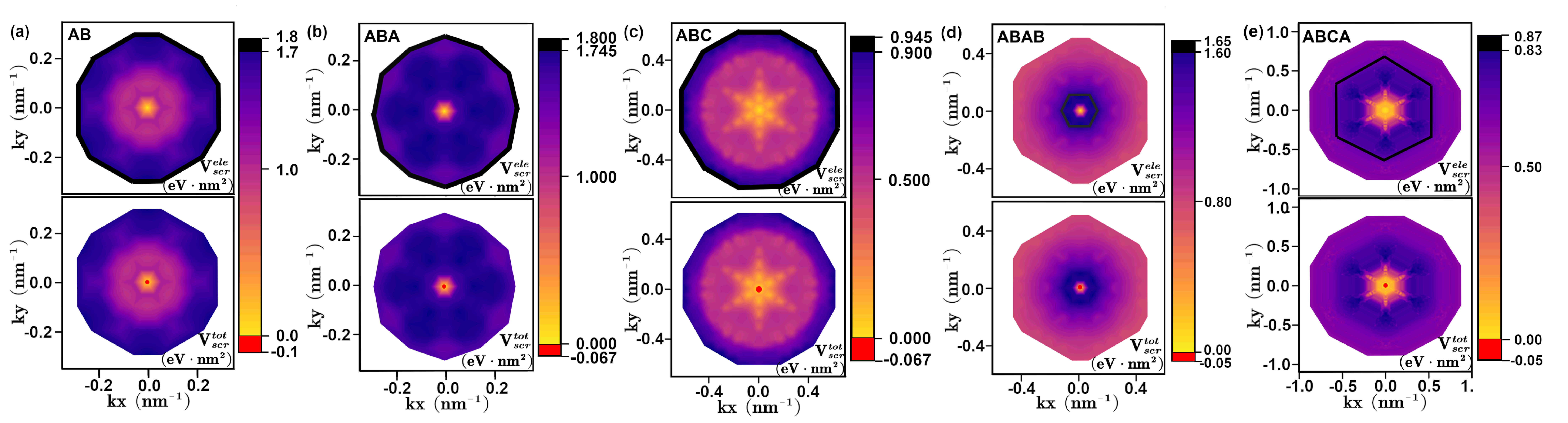}
  \caption{(Top panel) Momentum space screened Coulomb potential due to electron-electron interactions only, $V^{ele}_{scr}(\mathbf{q})$ in the five multilayer graphene stacks.\ (Bottom panel) The corresponding screened Coulomb potential $V^{tot}_{scr}(\mathbf{q})$ due to both electron-electron interactions and electron-phonon coupling.\ The interlayer potential is set to $\rm \Delta_1 = 50$ meV in all cases.}
  \label{fig:3}
\end{figure*}

The diagrams that describe the screened interaction are shown in Fig.~\ref{fig:2}(a).\ The straight blue lines stand for the long-range acoustic phonons coupling to the electrons, which contributes to the screened Coulomb interaction \cite{kohn1965new,cea2021coulomb}.\
We consider that longitudinal phonons and electrons couple through the deformation potential $V_D$ in the graphene system.\ This interaction can be described by a momentum and frequency-dependent potential ${V}^{{ph}}(\mathbf{q},{\omega})$ given by:
\begin{equation}
    V^{ph}(\mathbf{q},{\omega})=\frac{{ V}_{D}^2|\mathbf{q}|^2}{\rho(\omega^2-v_{s}^2|\mathbf{q}|^2)}
    \label{equ:3}
\end{equation}
where $\rho$ is the mass density, $v_{s}=\sqrt{(\lambda_{L}+2\mu_{L}) / \rho}$ is the velocity of sound and $\lambda_{L}$ and $\mu_{L}$ are the elastic Lam\'e coefficients \cite{michel2008theory}.\ In the following we take the zero-frequency limit of Eq.\ (\ref{equ:3}) which translates into a momentum independent electron-phonon interaction, given by:
\begin{equation}
    V^{ph}(\mathbf{q}) = -\frac{V_{D}^{2}}{\lambda_{L} + 2\mu_{L}}
    \label{equ:4}
\end{equation}
We set $\lambda_{L}+2\mu_{L} = 2\times10^3$ eV $\cdot$ nm$^{-2}$~\cite{cea2021coulomb} and consider that the deformation potential is given by $V_{D}=20/\sqrt{N}$, where $N$ is the number of layers of the multilayer graphene, for details see Appendix \ref{append_defor}. 
These considerations lead to $V^{ph} = -0.2/ \sqrt{N}$ eV $\cdot$ nm$^2$ for any momentum $\mathbf{k}$, although for computational simplicity we set an ultraviolet cut-off $\mathbf{q}_{c}$, defined by $\abs{\mathbf{q}_{c}}=\abs{\mathbf{q}_{K}}/10$, where $\mathbf{q}_{K}$ is the Dirac point momentum.\ 

We apply the RPA \cite{verma2012increasing,chen2017random,anderson1958random} to compute the renormalization of the Coulomb potential, as shown diagrammatically in Fig.~\ref{fig:2}(b).\ The total screened Coulomb potential $V^{tot}_{scr}(\mathbf{q})$ due to both electron-hole excitations and acoustic phonons coupling with electrons can be written as:
\begin{equation}
    V^{tot}_{scr}(\mathbf{q})=\frac{V_{C}(\mathbf{q})}{1 - V_{C}(\mathbf{q})\Pi^{tot}(\mathbf{q})}
    \label{equ:5}
\end{equation}
where the polarizability $\Pi^{tot}(\mathbf{q})$ including electron-acoustic phonon coupling can be written as the renormalization of the bare electronic polarizability $\Pi^{ele}(\mathbf{q})$:
\begin{equation}
    \Pi^{tot}(\mathbf{q}) = \frac{\Pi^{ele}(\mathbf{q})}{1 - V^{ph}(\mathbf{q})\Pi^{ele}(\mathbf{q})}
    \label{equ:6}
\end{equation}
where the bare electronic polarizability $\Pi^{ele}(\mathbf{q})$, depicted as a black oval in Fig.~\ref{fig:2}, is described as \cite{hwang2007dielectric, giuliani2005quantum, richardson1997effective}:
\begin{equation}
\label{equ:7}
\begin{aligned}
\Pi^{ele} \left( \mathbf{q} \right) \,\,=\,\,\frac{g_s}{N_C A_C}\sum_{\mathbf{k}, n,m}\frac{f(\xi_{{n},\mathbf{k}})-f(\xi _{{m},\mathbf{k}+\mathbf{q}})}{\varepsilon _{{ n},\mathbf{k}}-\varepsilon _{{ m},\mathbf{k}+\mathbf{q}}}\\
\times \left| \left< \psi _{{m},\mathbf{k}+\mathbf{q}}|\psi _{{n},\mathbf{k}} \right> \right|^2
\end{aligned}
\end{equation}
Here $g_s=2$ stands for the spin degeneracy, $N_C$ is the number of unit cells in the system, which we set up to $N_C=3\times 10^6$ after analyzing convergence of the calculations, $A_C=\frac{\sqrt{3}a^2}{2}$ is the area of each unit cell, $f(\xi_{{n},\mathbf{k}})=1/(1+{\rm e}^{\xi_{{n},\mathbf{k}}/{k_B T}})$ is the Fermi-Dirac distribution, with $\xi_{{ n},\mathbf{k}}=\varepsilon_{{n},\mathbf{k}}-\mu$, being $\mu$ the Fermi energy.\ $\ket{\psi_{n,\mathbf{k}}}$ and $\varepsilon _{{ n},\mathbf{k}}$ are the wavefunction, $k_B$ is the Boltzmann constant, and the corresponding $n$-th band energy with momentum $\mathbf{k}$.

The total screened potential $V^{tot}_{scr}(\mathbf{q})$ in Eq. (\ref{equ:5}) includes screening due to both electron-electron interaction and electron-phonon interactions. To investigate the effect of the electron-phonon interaction on the superconductivity, we also study the potential due to only electron-electron interactions, given by:
\begin{equation}
    V^{ele}_{scr}(\mathbf{q})=\frac{V_{C}(\mathbf{q})}{1 - V_{C}(\mathbf{q})\Pi^{ele}(\mathbf{q})}
    \label{equ:8}
\end{equation}

In the following, we compare the results considering the two possibilities for the screened potential:\ either it comes from electron-electron interactions alone, Eq.~(\ref{equ:8}), or it comes from these interactions plus electron-phonon dressing, Eq.~(\ref{equ:5}).\ In  Fig.~\ref{fig:3} we show the screened potentials in both cases for the five different multilayer graphene stacks.\ The potentials are plotted near the center of the first BZ.\ The main feature of the potentials is that the minimal values are at $\mathbf{q}=0$, and increase with $\mathbf{q}$.\ The inclusion of electron-phonon coupling leads to small attractive regions near $\mathbf{q}=0$, which do not significantly change the real-space screened potential, details in Appendix~\ref{append_Vx}.\

\subsection{\label{sec:2C}Superconductivity: Kohn-Luttinger-like mechanism}

In the Kohn-Luttinger (KL) mechanism the superconducting instability is driven by the screening of the interacting potential, which at long-range manifest an oscillatory behaviour giving room for the appearance of attractive regions that promote the formation of Cooper pairs \cite{kohn1965new}.\ At an effective level the interacting potential is a combination of all possible interactions present in the system, such as the direct Coulomb, the exchange or the electron-phonon interactions, among others.\
In this work we consider the direct Coulomb interaction to infinite order via the RPA, along with the electron-phonon dressing, while we omit the exchange potential.\ This approximation can be safely done since in the case of graphene-based systems the number of flavours is up to 4, implying that the direct diagrams are four-fold degenerate in contrast to the exchange interactions, see Refs.\ \cite{Cea2022SCKLRTG,PhysRevB.107.L161106}.

\begin{figure*}[htbp]
   \centering
   \includegraphics[width=0.9\textwidth]{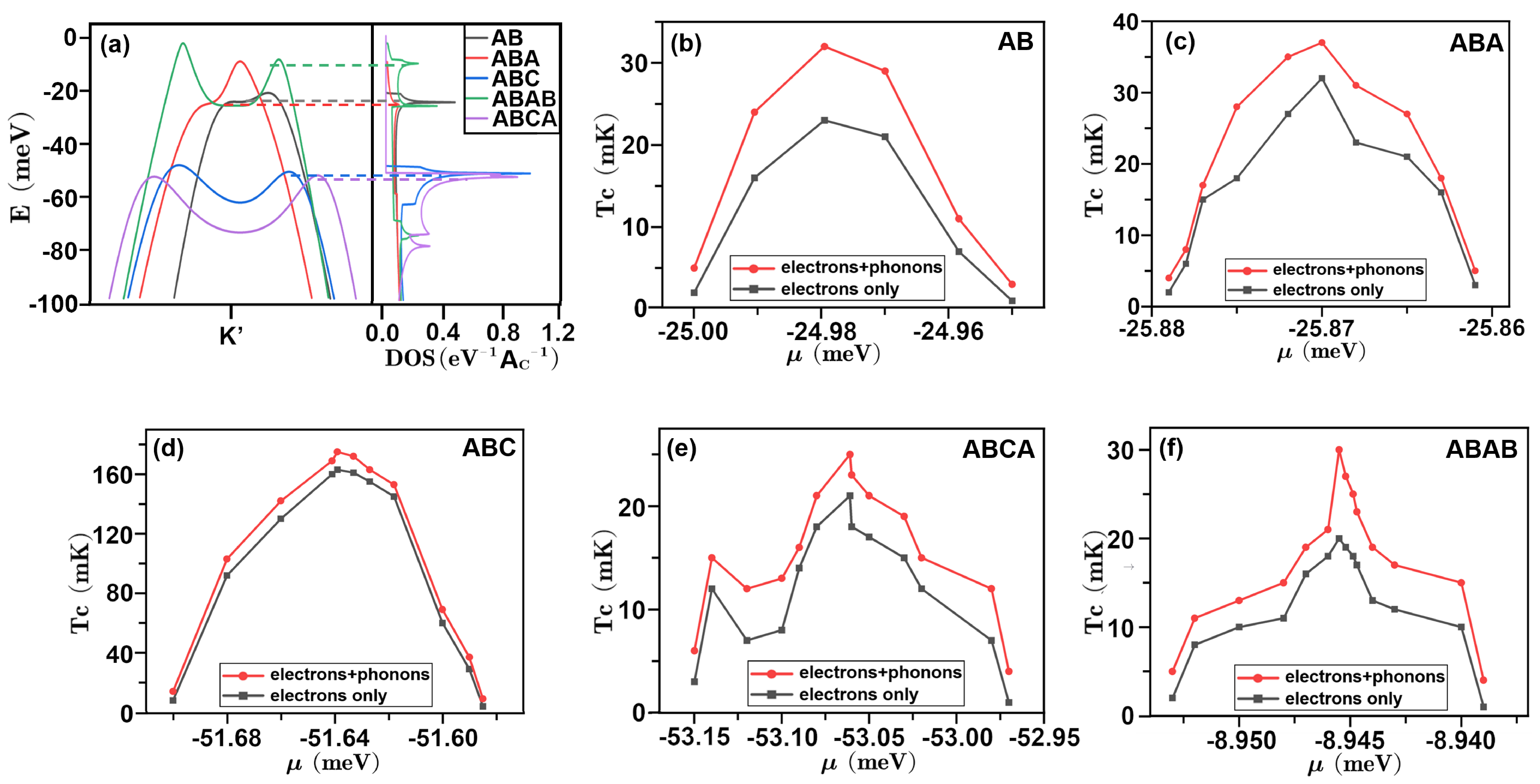}
   \caption{\label{fig:4}(a) Band structures and DOS near the CNP in multilayer graphene.\ (b-f) Superconducting critical temperatures versus Fermi energy near the VHS in five different multilayer graphene stacks, due to electron interactions only or taking into account also the effects of electron-acoustic phonon coupling.}
\end{figure*}

The KL-like approach to the problem leads to a self-consistent linearized gap equation  \cite{rajagopal1991linearized} given by:
\begin{equation}
\label{equ:9}
\begin{aligned}  
      &&\Delta^{ij}(\mathbf{k}) &= \sum_{i',j'}\sum_{\mathbf{k}',\omega}\frac{-k_B T}{N_C A_C}{V_{scr}}(\mathbf{k}-\mathbf{k}') &\\ 
      &&     &     \times G^{ii'}(\mathbf{k}',{\rm i\hbar} \omega)G^{jj'}(-\mathbf{k}',{\rm -i\hbar} \omega) \Delta^{i'j'}(\mathbf{k}') 
 \end{aligned}    
\end{equation}
where $G^{ii^{\prime}}\left(\mathbf{k},\pm\mathrm{i}\hbar\omega\right) = \sum_{m}\frac{\psi_{m,\mathbf{k}}^{i}\psi_{m,\mathbf{k}}^{i^{\prime}}}{\mathrm{i}\hbar\omega \mp \xi_{m, \mathbf{k}}}$ is the Green function with $i, i'$ labelling the atom position in the unit cell lattice while $\omega$ stands for the Matsubara frequencies. The $V_{scr}$ can be the total screened Coulomb potential due to electronic interactions plus electron-acoustic phonon coupling from Eq.~(\ref{equ:5}) or the purely electronically screened Coulomb potential from Eq.~(\ref{equ:8}).\ Equation (\ref{equ:9}) can be rewritten by applying the Matsubara sum \cite{Nieto_1995,guerout2014derivation}:
\begin{equation}
    \Delta^{mm'}(\mathbf{k})=\sum_{\mathbf{k}',n,n'}\Gamma_{mm'nn'}(\mathbf{k},\mathbf{k}')\Delta^{nn'}(\mathbf{k}')
\label{equ:10}
\end{equation}
where $\Delta^{mm'}(\mathbf{k})$ describes the amplitude of the Cooper pairing between bands $m$ and $m'$. The Hermitian kernel $\Gamma_{mm'nn'}(\mathbf{k},\mathbf{k}')$ can be written as \cite{PhysRevB.107.L161106}:
\begin{equation}
  \begin{aligned}
  &&\Gamma_{mm'nn'}(\mathbf{k},\mathbf{k}') = -\frac{1}{N_C A_C} {V_{scr}}(\mathbf{k}-\mathbf{k}') \\
  && \times \left< \psi_{{ m},\mathbf{k}} | \psi_{{ n},\mathbf{k}'} \right> \left< \psi_{{n'},\mathbf{k}} | \psi_{{m'},\mathbf{k}'} \right> \\
  && \times \sqrt{\frac{f(-\xi_{{ m'},\mathbf{k}})-f(\xi_{{m},\mathbf{k}})}{\xi_{{m'},\mathbf{k}}+\xi_{{m},\mathbf{k}}}} \sqrt{\frac{f(-\xi_{{n'},\mathbf{k}'})-f(\xi_{{n},\mathbf{k}'})}{\xi_{{n'},\mathbf{k}'}+\xi_{{ n},\mathbf{k}'}}}
  \end{aligned}
\label{equ:11}
\end{equation}

After these transformations the gap equation in Eq.\ (\ref{equ:10}) is directly solved when the largest eigenvalue of the kernel reaches a value of one, establishing the onset for the superconducting phase as a function of the temperature and Fermi energy.\ To reduce the computational complexity of the eigenvalue problem we impose an energy cut-off for states that form the kernel.\ This means that we only consider states with an energy close to the Fermi level, i.e.\ $\abs{\xi_{m,\vb{k}}} \leq \varepsilon_{c}$.\ We set $\varepsilon_{c}=3$ meV, which is large enough for the convergence.
\section{Results}
\label{result}

\subsection{Band structures and density of states}
 
We first calculate the band structures and density of states (DOS) of the graphene multilayers.\ It is well known that the external perpendicular electric field opens a gap at charge neutrality (CNP) and quenches the band borders of the low energy bands \cite{zhang2009direct,mccann2013electronic}, enhancing the density near the CNP and leading to van Hove singularities (VHS) in multilayer graphene systems.\ We set the interlayer potential to $\Delta_1 = 50$ meV \textcolor{black}{(results of different $\Delta_1$ 
in Appendix~\ref{append_Delta1}), and mainly focus on the effects of different stackings and different number of layers.} \ In Fig.~\ref{fig:4}(a), we show the lowest energy band in the hole side near the CNP of multilayer graphene.\ ``Mexican hat" profiles appear at the band edges due to the presence of the perpendicular electric field, making the density diverge logarithmically.\ The DOS of these graphene stacks behave differently under the same electric field, which partly explains the differences in the screened Coulomb interaction seen in Fig.~\ref{fig:3}.\ Note that the ABC trilayer graphene possesses the maximum value of the density at the VHS in Fig. \ref{fig:4}(a).\ The strength of the electric field changes the DOS in the graphene stacks as well, which is beyond the scope of this paper \cite{CWSS22b}.\ We mainly focus on the effect of the number of layers and their stacking on the superconductivity of non-twisted multilayer graphene.\ 

\subsection{Critical temperatures\label{Tc}}
The superconducting critical temperatures around the VHS are obtained by solving the Eq.\ (\ref{equ:10}).\ In Fig.~\ref{fig:4}(b-f), we show the critical temperatures $T_c$ as a function of the Fermi energy $\mu$ for the five graphene stacks.\ The critical temperatures strongly depend on electron filling and reach the maximum values near the energy at which the DOS diverge.\ We present the results obtained from the purely electronic interaction and compare them with the results including electron-phonon coupling.\ The critical temperatures $T_c$ increase only a few milli-Kelvin when including the electron-phonon coupling, which means that phonons have negligible contribution to the superconductivity.\ 



Recent experiments on superconductivity in AB- and ABC-stacked graphene \cite{Zhou2021SuperRTG,Zhou2022SCBG} measure $T^{AB}_{exp} = 26$ mK and $T^{ABC}_{exp} = 106$ mK,  respectively.\ We first apply our mechanism to the $T_c$ calculation in these two systems \cite{talantsev2020classifying}, with potential bias $\Delta_1 = 50$ meV.\ As shown in Fig.~\ref{fig:4}(b, d), we obtain $ T^{AB}_c \approx 23$ mK and $T^{ABC}_c \approx 160$ mK. Our calculated results match very well with the experimental values of $T_c$.\ This agreement with the experiments shows the prospective reliability of our mechanism.

\begin{figure*}[htbp]
   \centering
   \includegraphics[width=0.9\textwidth]{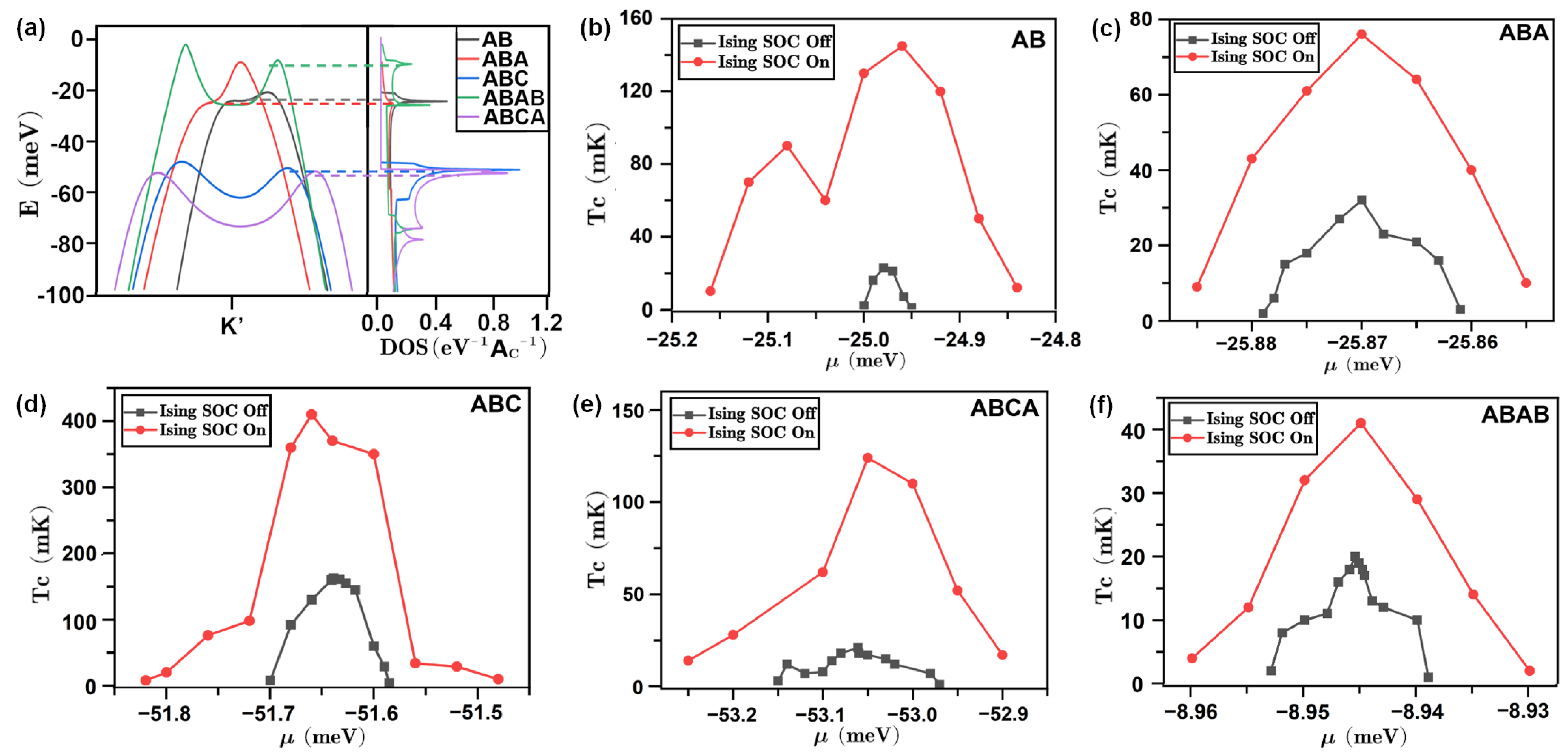}
   \caption{\label{fig:Ising}\textcolor{black}{(a) Bands and DOS near VHS in multilayer graphene, with displacement field $\Delta_1 = 50$ meV. (b)-(f) Critical temperature of multilayer graphene with and without Ising SOC, with fermi energy near VHS.}}
\end{figure*}

Among these five stacks, rhombohedral ABC trilayer graphene has the highest $T_c\sim100$ mK, while all other stacks (surprisingly even rhombohedral ABCA tetralayer graphene) have a $T_c$ lower than 40 mK.\ This has the important implication that $T_c$ is not strongly correlated with the number of layers.\ Other features, such as the form of the order parameters, are much more relevant for superconductivity, as discussed below.\ These results provide guidance to future search for superconductivity in non-twisted multilayer graphene.

\subsection{Ising superconductivity}
 
Recent experimental findings \cite{zhang2022spin,holleis2023ising} indicate that, in the presence of Ising SOC, AB bilayer graphene is an Ising superconductor, which means that it has spin-valley locked Cooper pairs of type e.g.\ $\ket{\vb{K}^{+},\uparrow; \vb{K}^{-},\downarrow}$, with only one spin orientation per valley.\ To a first approximation, the effect of Ising SOC is to raise two flavours with respect to the other two, separating them by an energy $\lambda_I\sim$ 1-2 meV. This breaks the equivalence between Cooper pairs $\ket{\vb{K}^{+},\uparrow; \vb{K}^{-},\downarrow}$ and $\ket{\vb{K}^{+},\downarrow; \vb{K}^{-},\uparrow}$ and splits the four-fold degenerate VHS into two VHS which are two-fold degenerate, which leads to two superconducting sleeves in the phase diagram of bilayer graphene \cite{holleis2023ising}.\ In our framework, the energy splitting results in a modified susceptibility.\ Here we focus only on the VHS that appears at lower Fermi energy after the splitting, dubbed ``SC2'' in Ref.\ \cite{holleis2023ising}, which shows stronger superconductivity.\ The largest contribution to the susceptibility comes from this VHS, while the one at higher Fermi energy gives a small contribution, which we neglect here.\ Therefore, we approximate the spin-valley locked susceptibility by putting a factor of $g_s=1$ instead of 2 in Eq. (\ref{equ:7}).\ The quality of this approximation is discussed in Ref.\ \cite{PhysRevB.107.L161106}.\ Compared to a calculation that includes both contributions, we expect that the results presented here slightly overestimate $T_c$.\ However, they should give the right order of magnitude and allow us to identify trends.

 We present the critical temperatures in different multilayer graphene stacks with spin-valley locking \textcolor{black}{in Fig.~\ref{fig:Ising},}.\ AB bilayer graphene shows the largest enhancement, with maximum $T_c$ increasing by a factor of 6, from 23 mK to 143 mK due to electron-electron interactions only. \textcolor{black}{The superconductivity also occurs within a 7 times wider Fermi energy range, from 0.05 meV to 0.35 meV. These two features are in good agreement with recent experimental observations, in which factors of 10 times enhancement and 8 times wider energy range have been observed \cite{zhang2022spin,holleis2023ising}.}\ In ABA and ABAB stacks, for which the nearly constant sign of the order parameters suggests that intervalley (short-range) interactions are dominant, Ising SOC leads to more modest increments.\ This is reasonable because SOC does not change much the screened potential at the large $\mathbf{q}$ that corresponds to intervalley scatterings.\ Notably, rhombohedral trilayer (ABC) graphene experiences the smallest increment in $T_c$, suggesting that intervalley interactions are also important in this material.\ Interestingly, the ABCA stack also has a significant enhancement of superconductivity due to Ising SOC, \textcolor{black}{which could be induced by placing monolayer tungsten diselenide (WSe$_2$) on graphene stackings \cite{zhang2022spin,holleis2023ising}.} Therefore, stacking WSe$_2$ on this material should help to find superconductivity experimentally.\\

\subsection{Order parameters}

\begin{figure*}[htbp]
   \centering
   \includegraphics[width=\textwidth]{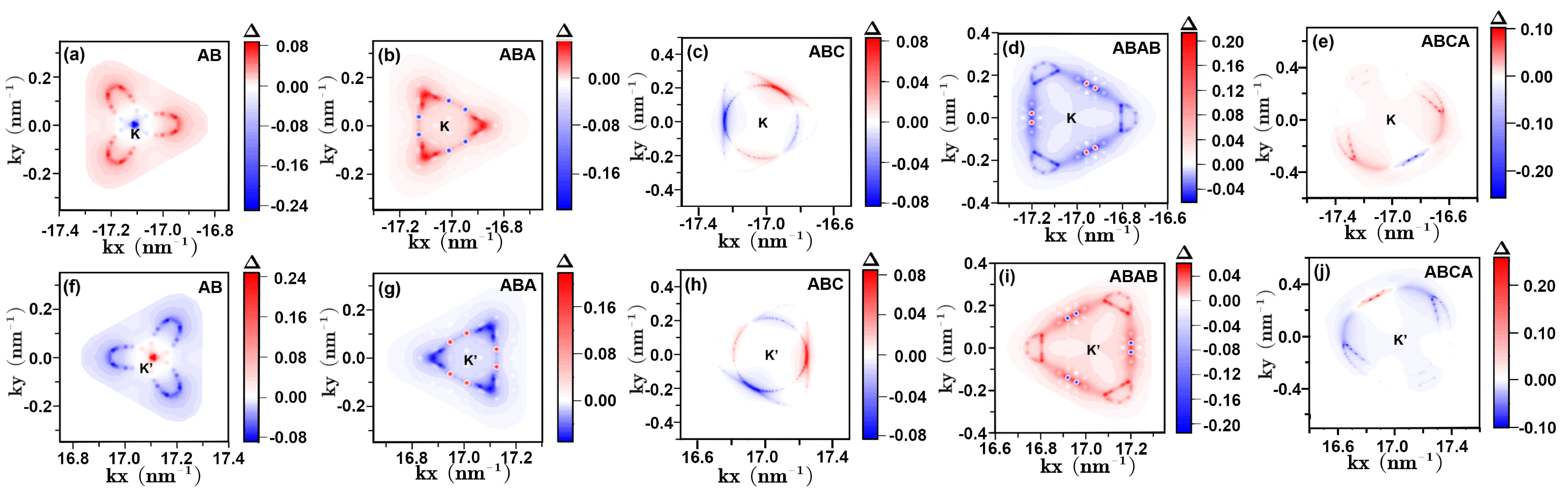}
   \caption{\label{fig:5}(a-e) Superconducting order parameters near K valley in the first BZ of multilayer graphene with interlayer potential $\Delta = 50$ meV and maximum critical temperatures $T_c$.\ (f-j) Order parameters near K' valley in the first BZ.}
\end{figure*}

Figure \ref{fig:5}(a-j) displays the superconducting order parameters $\Delta(\mathbf{k})$ in the K and K' valleys of the five graphene stacks, in which only the electron-electron interaction is considered.\ 
The OPs reveal a wealth of information \textcolor{black}{near the VHS.}\ For example, in Bernal-stacked (AB, ABA and ABAB) graphene, $\Delta(\mathbf{k})$ preserves $C_3$ symmetry, and has intravalley extended $s$-wave symmetry.\ In contrast, in rhombohedral-stacked (ABC and ABCA) graphene, $\Delta(\mathbf{k})$ breaks $C_3$ symmetry and has two-fold degenerate eigenvalues and $p$-wave-like symmetry.\ This is a significant difference between Bernal- and rhombohedral-stacked graphene.\ Some important similarities are also found in all order parameters.\ First, all the stacks show sign change and nodes within each valley. Second, all stacks change sign between valleys K and K'.\ As the total electron wave function must be anti-symmetric, this implies  that all these graphene systems are valley-singlet, spin-triplet superconductors. \textcolor{black}{Moreover, we investigate OPs with varying $\mu$ in the Appendix~\ref{append_OP}, suggesting that the stability of the order parameters near VHS.}

Once again, ABC-stacked graphene stands out due to its extended and balanced intensity stripes of opposite signs, which induce strong superconductivity.\ The nearly annular shape of the Fermi surface of ABC trilayer graphene can lead to nesting, which is expected to be beneficial for superconductivity driven by Coulomb interactions~\cite{PhysRevLett.127.247001}.\ This also helps to explain why the ABC stack has higher $T_c$ than the other four stacks.\ Finally, we note that the OPs reported here for AB bilayer and ABC trilayer graphene are in excellent agreement with those found in Ref.\ \cite{PhysRevB.107.L161106} within a continuum model.

\textcolor{black}{Some previous theory works~\cite{CWSS22,CWSS22b,Chou2021AcousticRTG} have proposed a direct electron-acoustic phonon interaction as responsible for superconductivity in non-twisted multilayer graphene.\ We believe that our purely electronic theory is much more likely to hold, for the following reasons:\ i) Our KL-like approach captures the enhancement of superconductivity due to Ising SOC in excellent agreement with experiments \cite{zhang2022spin,holleis2023ising}, both in terms of the increment in superconducting $T_c$ as well as the increment of the filling range over which superconductivity occurs, see Fig.\ 5 and discussion below.\ Moreover, the theory captures this enhancement in a natural way:\ the flavour polarization due to Ising SOC increases the magnitude of the screened potential, and it does so without changing its crucial unconventional shape \cite{MC13} (the potential is minimum at k=0 and grows with momentum).\ Such change in the screened potential directly leads to stronger superconductivity in our formalism.\ This is in contrast to theories in which new physical processes are invoked to explain the enhancement \cite{chou2022enhanced}.\ ii) The KL-like theory has no free parameters, in contrast to other theories, both electronic and conventional.\ iii) Our mechanism, which involves only electrons interacting, is simpler than the one that involves interactions between two types of species, electrons and phonons.\ iiii) The KL-like theory leads to critical temperatures in very good agreement with experiments, e.g. $T_c\sim25$mK in bilayer graphene.\ In contrast, phonon-mediated theories lead to $T_c\sim 1$K, unless a detrimental effect of Coulomb interactions is assumed, something that, as we show here, need not be the case.}

\section{Discussion}

The discovery of superconductivity in non-twisted systems without a moiré pattern, for example, AB bilayer and ABC trilayer graphene \cite{Zhou2021SuperRTG,Zhou2022SCBG,zhang2022spin}, brings forth a new stage in the field of graphene superconductors.\ With the picture of twisted graphene superconductors growing increasingly complex, reaching first an understanding of their non-twisted counterparts is a promising route to advance the field.\ 
Here, we have established that the screened Coulomb interaction alone explains superconductivity in AB bilayer and ABC trilayer graphene, see also Refs.\ \cite{PhysRevB.107.L161106,Cea2022SCKLRTG,PhysRevLett.130.146001,Lu2022RG_RTG,PhysRevLett.127.247001,Wagner2023Bernal,C22,Dai2022QMonteCarlo,DHZLM21}.\ Furthermore, we predict that the Coulomb interaction leads to superconductivity in the ABA trilayer and the ABAB and ABCA tetralayers, non yet found in experiments.\ Note that the Kohn-Luttinger-like framework has no free parameters fitted to match experimental results, and yet the calculated superconductivity match well the reported experimental results.\ We also dress the Coulomb interaction by acoustic phonons.\ Comparing the critical temperatures obtained with and without phonon dressing, we find that phonons play a secondary role in all cases.\ The results presented here suggest that the strength of superconductivity is not linearly correlated with the number of layers, but rather with more subtle features such as the characteristics of the Fermi surface \cite{PhysRevLett.127.247001}.

We show that Ising SOC promotes superconductivity in multilayer graphene, especially in the AB bilayer and ABCA tetralayer graphene.\ The RPA, KL-like mechanism for superconductivity offers a simple explanation for the enhancement of superconductivity due to Ising SOC \cite{zhang2022spin,holleis2023ising}.\ We note  here two key aspects of the mechanism \cite{PhysRevB.107.L161106}:\ Firstly, the screened potential is minimum at $\mathbf{q}
=\mathbf{0}$ and grows with $\mathbf{q}$;\ Secondly, such unconventional dependence allows order parameters with sign changes \cite{MC13}.\ Spin-valley locking reduces the screening.\ The screened potential retains its peculiar momentum dependence, but increases in magnitude.\ This reinforces interactions between states with different signs within the OP, which strengthen superconductivity.\ In fact, the reduced screening also reinforces interactions between states with the same sign, which are detrimental for superconductivity, but the former interactions prevail.\ Although somewhat paradoxical, this suggests that, once there is a repulsive potential as pairing glue, \textcolor{black}{such that its shape favours superconductivity}, the more repulsive the potential the better for superconductivity.\ Ising SOC also leads to spin-valley locked order parameters in these materials \cite{PhysRevB.107.L161106}, characteristic of Ising superconductivity.\ Such a state cannot be described as singlet or triplet.

The order parameters provide deeper insights into the nature of superconductivity.\ All graphene superconductors studied here, both Bernal-stacked and rhombohedral-stacked, share the following features: i) The OPs change sign between valleys.\ Since the electron wave function is anti-symmetric, this means that the pairs are spin-triplets.\ This also implies that short-range disorder is pair-breaking.\ 
ii) The OPs have changes in signs and nodes within each valley.\ As a consequence, the long-range disorder is also pair-breaking.\ Such order parameters are favoured by the peculiar form of the screened Coulomb potential \cite{MC13}.\ These sign changes are features of superconductors with weak coupling, in which the symmetry is $p$- or $f$-wave \cite{Cea2022SCKLRTG}.\ However, there are also important differences:\ i) The OPs of Bernal stacks are $C_3$-symmetric, and display extended $s$-wave symmetry.\ This is consistent with the recent experiment~\cite{Lin23Spontaneous} on AB bilayer graphene, which reveals spontaneous momentum polarization ($C_3$-symmetry breaking) all across its phase diagram, except in stripes at phase boundaries like the one in which superconductivity emerges, which preserve $C_3$ symmetry.\ In contrast, the OPs of rhombohedral stacks break $C_3$ and have $p$-wave-like symmetry.\ This suggests that momentum polarization also exists in the phase diagram of rhombohedral stacks, and even in the superconducting phase.\ ii) The OP of rhombohedral ABC graphene has balanced intensity stripes with opposite signs, which indicate that intravalley (long-range) scatterings boost the strong superconductivity of this material.\ iii) The OPs of Bernal ABA and ABAB graphene have a dominant sign within the valley with a few hotspots of opposite sign, which suggests that the intervalley (short-range) interaction is the main contribution to superconductivity.

\hspace{3pt}
\section{Acknowledgements}
We thank Pierre A. Pantaleon, Tommaso Cea and Yunhai Li for their useful discussions. This work was supported by the National Natural Science Foundation of China (Grant No. 12174291) and the National Key R\&D Program of China (Grant No. 2018YFA0305800). 
IMDEA Nanociencia acknowledges support from the ``Severo Ochoa" Programme for Centres of Excellence in R\&D (CEX2020-001039-S / AEI / 10.13039/501100011033). Z.Z. acknowledges support funding from the European Union's Horizon 2020 research and innovation programme under the Marie Skłodowska-Curie grant agreement No 101034431 and from the ``Severo Ochoa" Programme for Centres of Excellence in R\&D (CEX2020-001039-S / AEI / 10.13039/501100011033).
Numerical calculations presented in this paper have been performed on the supercomputing system in the Supercomputing Center of Wuhan University.

\appendix

\section{Dependence of superconductivity on the background dielectric constant $\epsilon$}
\label{append_eps}

\begin{figure}[H]
  \centering
  \includegraphics[width=0.5\textwidth]{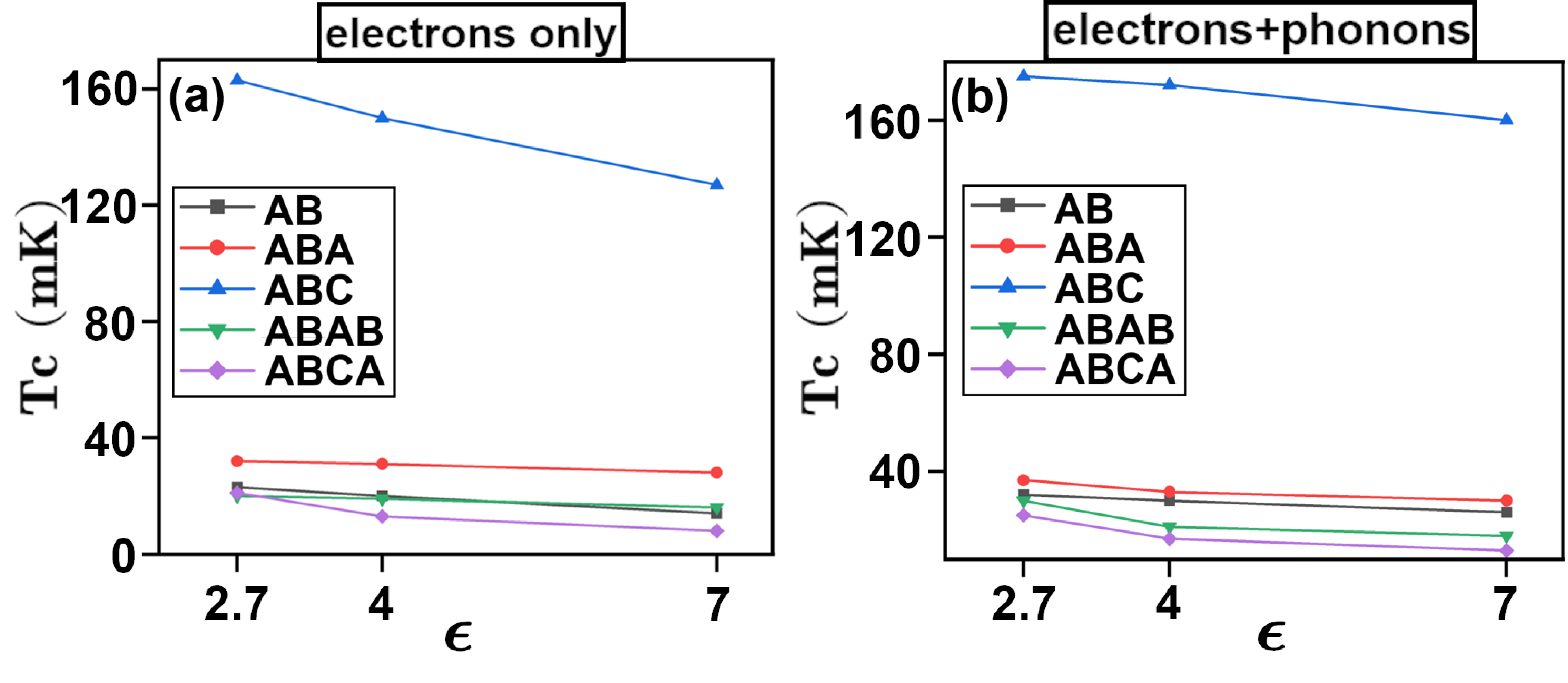}
  \caption{Superconducting critical temperatures of different graphene systems, as a function of the background dielectric constants $\epsilon$. (a) Due to electron-electron interactions only.\ (b) Considering also electron-acoustic phonon coupling.}
  \label{fig:7}
\end{figure}

We investigate here the effects of changing the dielectric constant due to hBN encapsulation, $\epsilon$, which is the least well determined parameter in our model.\ Figure~\ref{fig:7} shows the critical temperatures of the five multilayer graphene stacks for different values of $\epsilon$.\ The results show that increasing the dielectric constant has only a weak impact on superconductivity in every stack.\ This implies that our results are robust against changes in the dielectric constant.

\section{Coupling between charge oscillations and longitudinal acoustic phonons in multilayers}
\label{append_defor}
We analyze the coupling between charge oscillations and longitudinal accoustic phonons in multilayers.\ We show that in system with $N$ layers the deformation potential $V_D$ is renormalized to $V_D/\sqrt(N)$.\ Within layer $i$, we assume lattice displacements $\{ u_x^i , u_y^i \}$ lead to a strain tensor $u^i_{\alpha , \beta} = ( \partial_\alpha u^i_\beta + \partial_\beta u^i_\alpha ) /2 $, where $\alpha , \beta = x , y$.\ These strains lead to an electrostatic potential defined as:
\begin{align}
V^i ( \mathbf{r} ) &= V_D \sum_{\alpha = x , y} u^i_{\alpha , \alpha}
\label{potential}
\end{align}
This potential changes the energy of the layer by:
\begin{align}
\delta E^i &= \rho^i ( \mathbf{r} ) V^i ( \mathbf{r} )
\label{energy}
\end{align}
where $\rho^i( \mathbf{r} )$ describes the electronic charge in layer $i$. The displacements, $u^{i}_{LA , x} , u^i_{AO , y}$, induced by a longitudinal acoustic (LA) phonon of momentum $\{ k_x , k_y \}$, are such that the vectors $\vec{u}^i_{LA}$ and $\vec{k}$ are parallel.\ Then, quantizing Eq.\ (\ref{potential}) and Eq.\ (\ref{energy}), the coupling between charge fluctuations and LA phonons is:
\begin{align}
{\cal H}^i_{e-ph} &= \sum_{\mathbf{q}} \hat{\rho^i}_{\mathbf{q}} | \mathbf{q} |\hat{u}^i_{LA \mathbf{q}} = \nonumber \\
&= V_D \sum_{\mathbf{q}} \hat{\rho^i}_{\mathbf{q}} | \mathbf{q} | \sqrt{\frac{\hbar}{2 m \omega_{LA } ( \mathbf{q} )} }  \left( b^{i \dag}_{LA \mathbf{q}} + b^i_{LA \mathbf{q}} \right)
\end{align}
where $m$ is the mass density, and the phonon frequency is:
\begin{align}
\omega_{LA } ( \mathbf{q} ) &= v^L_s | \mathbf{q} |
\end{align}
where the longitudinal sound velocity is:
\begin{align}
v^L_s = \sqrt{\frac{\lambda + 2 \mu}{m}}
\end{align}
where $\lambda$ and $\mu$ are elastic Lamé coefficients.

We neglect the elastic interactions between layers in the multilayer.\ Then, the phonons of the layers are not coupled.\ The electron-phonon coupling of the whole system is:
\begin{align}
{\cal H}_{e-ph} &= V_D \sum_{\mathbf{q}} | \mathbf{q} |\sqrt{\frac{\hbar}{2 m \omega_{LA } ( \mathbf{q} )} } \sum_{i=1}^{i=N} 
 \hat{\rho^i}_{\mathbf{q}}    \left( b^{i \dag}_{LA \mathbf{q}} + b^i_{LA \mathbf{q}} \right)
\end{align}
We make a canonical transformation on the phonons, and define:
\begin{align}
b^{l \dag}_{LA \mathbf{q}} &= \frac{1}{\sqrt{N}} \sum_{i=1}^{i=N} e^{j\frac{2 \pi l i}{N}} b^{i \dag}_{LA \mathbf{q}}
\label{trans}
\end{align}
where $l = 0 , \cdots , N-1$ and j is the imaginary unit.\ Using this transformation, we can write:
\begin{align}
{\cal H}_{e-ph} &= V_D \sum_{\mathbf{q}} | \mathbf{q} | \sqrt{\frac{\hbar}{2 m \omega_{LA } ( \mathbf{q} )} }    \frac{b^{0 \dag}_{LA \mathbf{q}} + b^0_{LA \mathbf{q}}}{\sqrt{N}}\sum_{i=1}^{i=N} 
 \hat{\rho^i}_{\mathbf{q}} + \cdots
\end{align}
The operator $\hat{\rho}_{\mathbf{q}} = \sum_{i=1}^{i=N} 
 \hat{\rho^i}_{\mathbf{q}}$ describes a charge fluctuation of the entire multilayer.\ The phonons in Eq.\ (\ref{trans}) with $l \ne 0$ couple to charge fluctuations of individual layers which average to zero over the multilayer, so that they do not induce electrostatic potentials at distances larger than the width of the multilayer.\ Hence, the coupling of phonons to long range charge oscillations, averaging over all layers, can be written as:
 \begin{align}
{\cal H}_{e-ph} &\approx \frac{V_D}{\sqrt{N}} \sum_{\mathbf{q}} | \mathbf{q} | \sqrt{\frac{\hbar}{2 m \omega_{LA } ( \mathbf{q} )} } \left(   b^{0 \dag}_{LA \mathbf{q}} + b^0_{LA \mathbf{q}} \right)
 \hat{\rho}_{\mathbf{q}} 
\end{align}
\\

\section{Screened potential in real space}
\label{append_Vx}
\begin{figure}[H]
  \centering
  \includegraphics[width=0.5\textwidth]{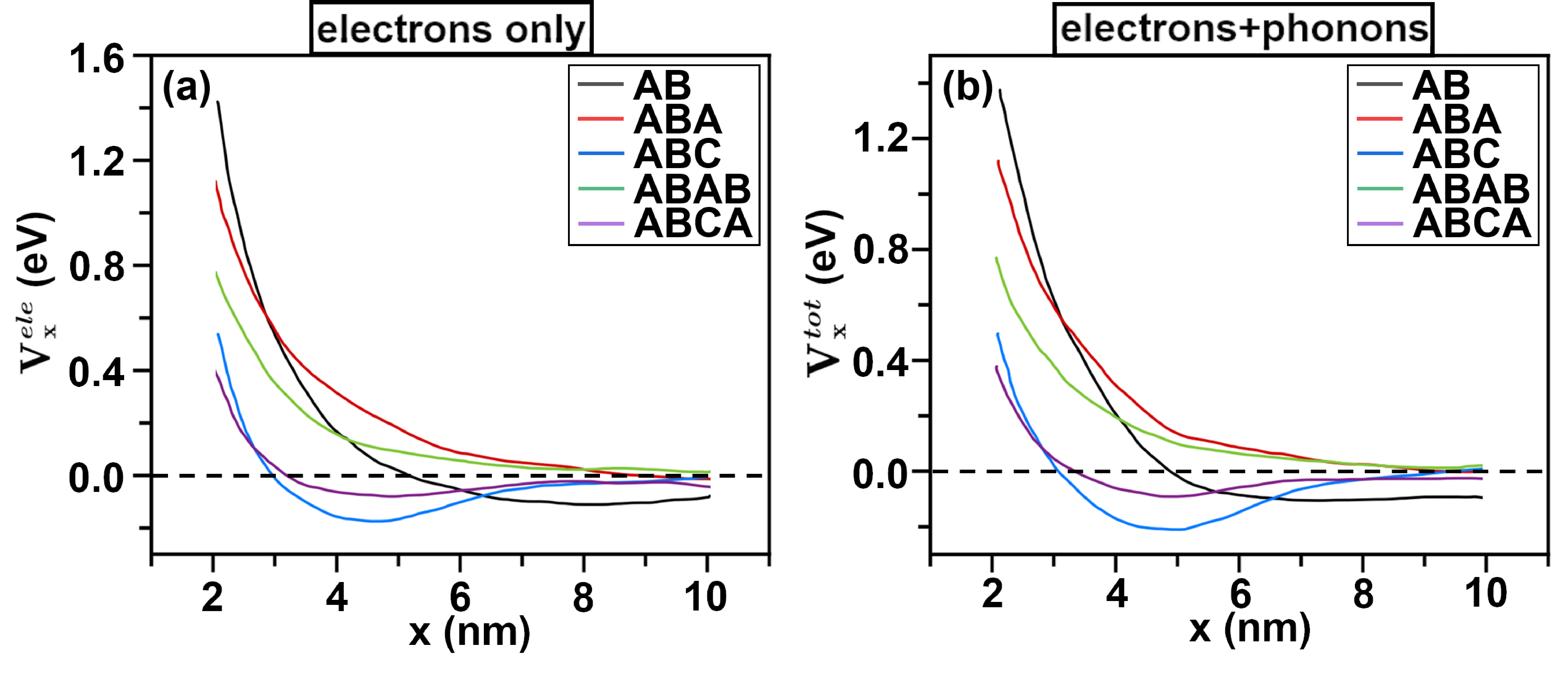}
  \caption{The real-space screened potential, as a function of x, in five graphene systems. (a) $V^{ele}_x$, screened potential with only electron-electron interactions. (b) $V^{tot}_x$, total screened potential including also the acoustic phonon-electron coupling. }
  \label{fig:6}
\end{figure}

In Fig.\ \ref{fig:3} of the main text we have presented the screened potentials of the five stacks in momentum space.\ It is also illustrative to compute their real space profiles, which are given by the Fourier transform of the momentum-space potentials:
\begin{equation}
    {V^{ele/tot}_{x}}(\mathbf{x})=\frac{A_{BZ}}{N_k}\sum_{\mathbf{k}}{V^{ele/tot}_{x}}(\mathbf{k}){\rm e}^{-{\rm i}\mathbf{k} \cdot \mathbf{x}}
    \label{SIequ:8}
\end{equation}
where $x$ is the distance, $A_{BZ}$ is the area of the BZ and $N_k$ is the total k points spread inside the BZ.

Figure \ref{fig:6}(a) shows the results of the real space screened potential in all stacks, due to electron-electron interactions only.\ All potentials, originally repulsive, oscillate and develop attractive minima at some distances, similar to the phenomenon of Friedel oscillations.\ These distances roughly correspond to the sizes of the formed Cooper pairs.\ It is worth noting that in rhombohedral (ABC and ABCA stacks) the minimum develops around 4 nm, in very good agreement with the result of Ref.\ \cite{Cea2022SCKLRTG}, while in Bernal stacks the minimum is shallower (except in the bilayer) and develops at larger distance $\sim10$ nm.\ As shown in Fig.\ \ref{fig:6}(b), including dressing by phonons barely modifies the real-space screened potential, which is consistent with the minor effect of phonons on the superconductivity.

\section{\textcolor{black}{Maximum critical temperatures at different displacement fields}}
\label{append_Delta1}

\begin{figure}[htbp]
  \centering
  \includegraphics[width=0.45\textwidth]{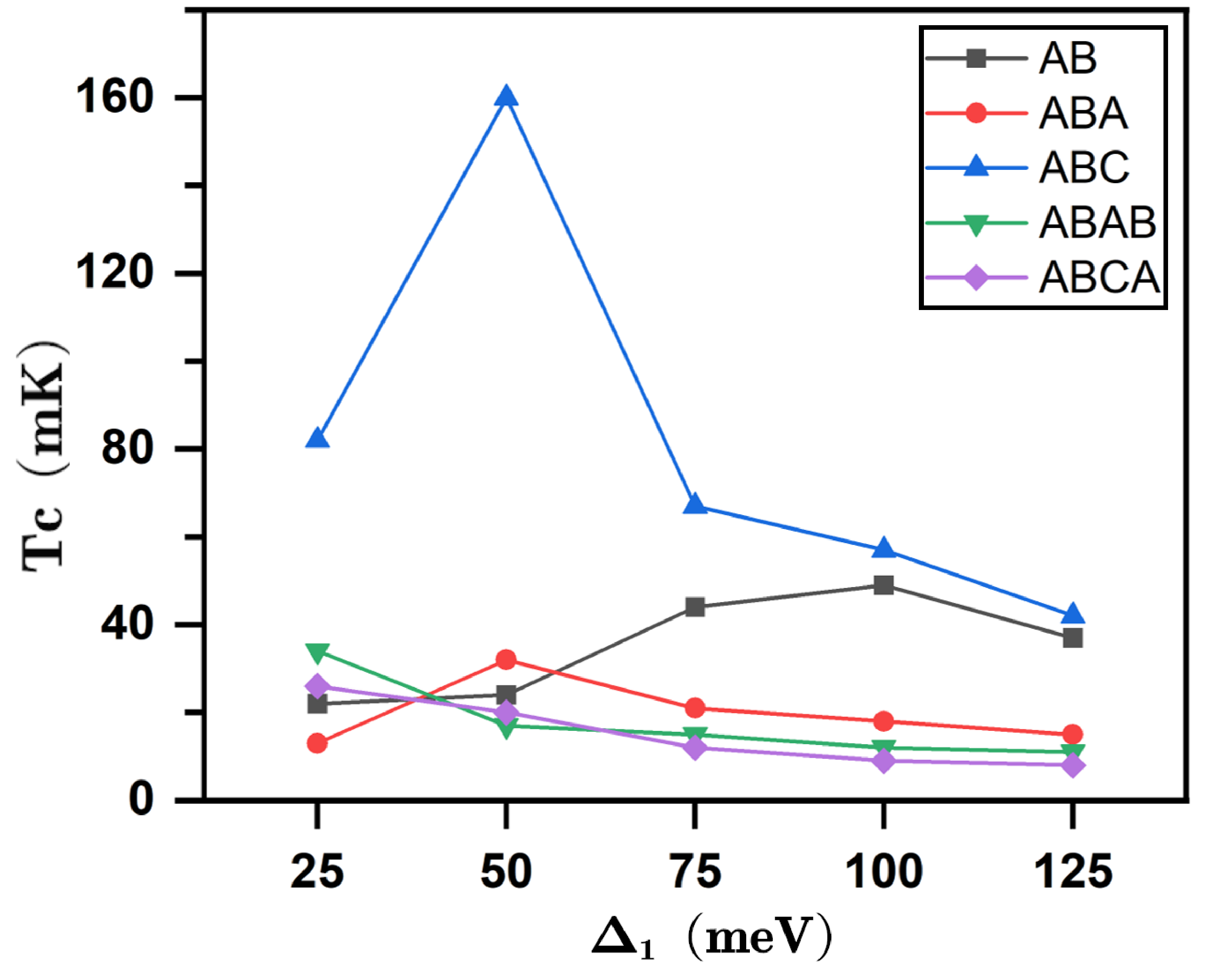}
  \caption{\textcolor{black}{The maximum critical temperatures in multilayer graphene with varying displacement fields $\Delta_1$.} }
  \label{fig:Delta1}
\end{figure}

\textcolor{black}{We compare the maximum critical temperatures in multilayer graphene with varying $\Delta_1 = 25,\ 50,\ 75,\ 100,\ 125$ meV, shown in Fig.\ \ref{fig:Delta1}. The background dielectric constants is set to $\epsilon = 2.7$. With varying displacement fields, ABC graphene has maximum $T_c \approx 160$ mK near $\Delta_1 = 50$ meV. For the AB stacking, the maximum $T_c$ is achieved with $\Delta_1 = 100$ meV. For the rest of stackings, the electric field has a minor change to the $T_c$. 
}

\section{\textcolor{black}{Stability of the order
parameters}}
\label{append_OP}

\begin{figure}[H]
  \centering
  \includegraphics[width=0.43\textwidth]{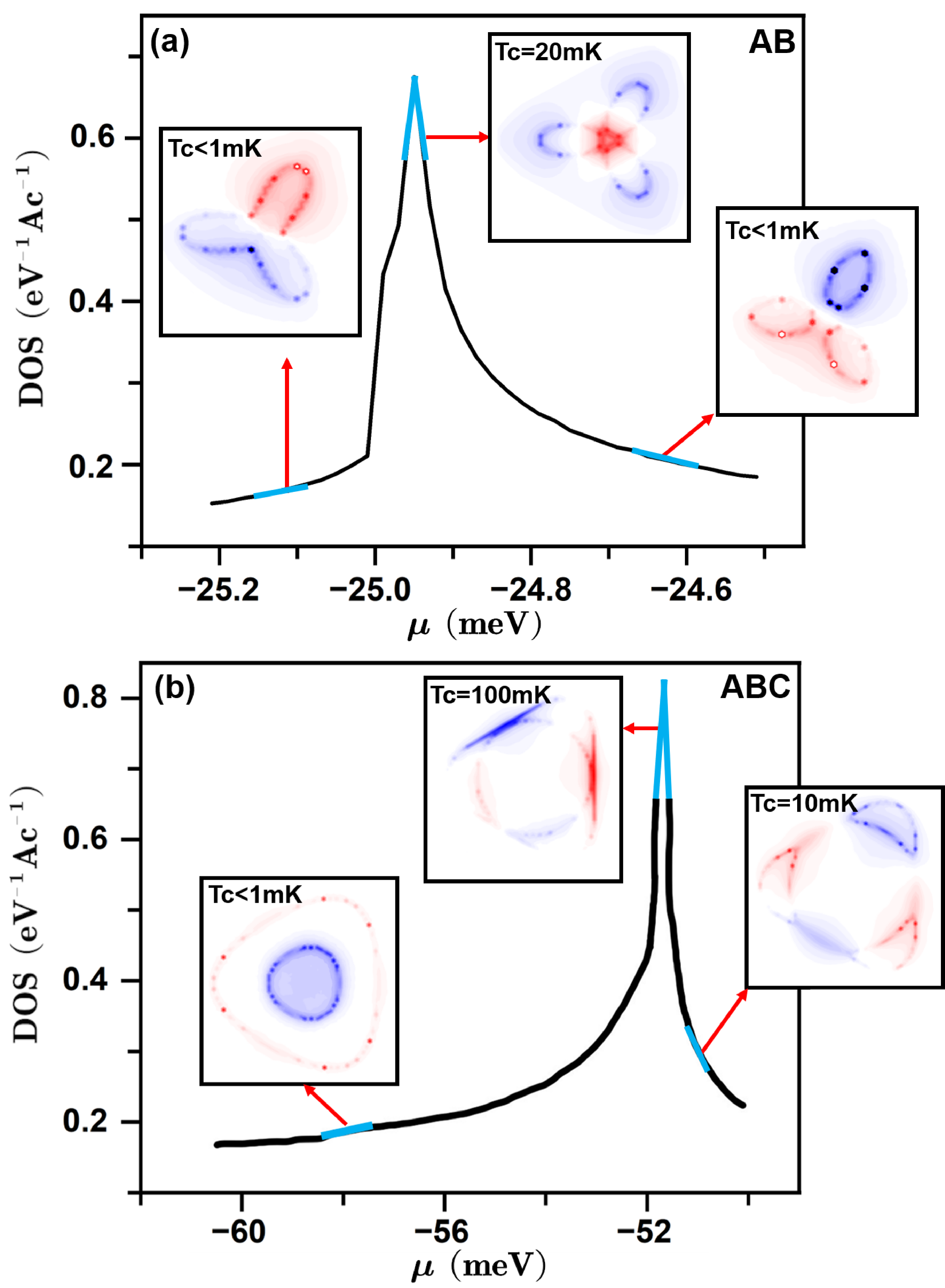}
  \caption{\textcolor{black}{DOS in AB and ABC graphene with varying Fermi energy near VHS, where the insets are the superconducting order parameters. The blue lines represent the range of $\mu$ where OPs maintain the same shapes. (a) OPs in AB graphene. (b) OPs in ABC graphene.} }
  \label{fig:OPs_mu}
\end{figure}

\textcolor{black}{We investigate OPs with varying $\mu$ near VHS. We take OPs in AB and ABC graphene with $\Delta_1 = 50$ meV as examples, which are representative of the Bernal and rhombohedral graphene. As shown in Fig.~\ref{fig:OPs_mu} (a) and (b), superconducting OPs will maintain $C_3$ symmetry in AB graphene with doping density close to VHS, while OPs of ABC graphene will break this symmetry. These findings suggest the stability of OPs in multilayer graphene near VHS. For Fermi energy not close to VHS, OPs show totally different shape and symmetry in both cases. Especially in ABC graphene, in the first inset of Fig.~\ref{fig:OPs_mu} (b) the OPs perform $C_3$ symmetry. These results are the same as the findings in Ref.~\cite{PhysRevLett.127.247001}. These phenomena are also detected in other Bernal and rhombohedral graphene stacks, which are not shown here.}

\providecommand{\noopsort}[1]{}\providecommand{\singleletter}[1]{#1}%

\end{document}